%% file: main.tex
\documentclass[conference]{IEEEtran}
\IEEEoverridecommandlockouts

\ifCLASSINFOpdf
   \usepackage[pdftex]{graphicx}
   \DeclareGraphicsExtensions{.pdf,.jpeg,.png}
\else
\fi
\usepackage{amsmath}
\usepackage{amssymb}
\usepackage{algorithmic}

\usepackage{array}
\usepackage{url}

\usepackage{multirow}
\usepackage{amsthm}
\usepackage{subcaption}
\theoremstyle{definition}

\usepackage{listings}
\usepackage{xcolor}
\definecolor{comments}{HTML}{868686}

\usepackage{mathtools}
\usepackage{dblfloatfix}
\lstdefinelanguage{riaps}{
 morekeywords={ uses, limits, use, for,  struct, cpu, component,max,mem, max_mem,cores,spc, max_spc,max_cpu, space, net,rate, kbps,ceil, burst, k, nodeLabel, nodeTemplate, artifact, device, located, at, mb, nodes,node, key,components,import,over,actor, colocate,timer,within,msec, pub, MB, rep,app, deploy, separate, on, copies, sec, message,'{','}'},
 keywordstyle=\color{blue}\bfseries,
    basicstyle=\scriptsize\ttfamily,
 morecomment=[l]{//}, 
 morecomment=[s]{/*}{*/},
 morecomment=[l]{\#},
 morestring=[b]",
    basicstyle=\scriptsize\ttfamily,%
 commentstyle=\color{comments}\ttfamily,
numbers=none,
    numberstyle=\scriptsize,
    stepnumber=1,
    numbersep=8pt,
breaklines=true,
    frame=tb,
 tabsize=2}
 
 \lstdefinelanguage{bash}{
 basicstyle=\ttfamily,
  showstringspaces=false,
  commentstyle=\color{red},
  keywordstyle=\color{blue},
  frame=tb
 }

\begin{document}
%
\title{An Automated Deployment and Testing Framework for Resilient Distributed Smart Grid Applications \\
\thanks{978-1-6654-8356-8/22/\$31.00 \copyright 2022 IEEE}
}



%

\author{\IEEEauthorblockN{Purboday Ghosh \IEEEauthorrefmark{1}, Hao Tu \IEEEauthorrefmark{2}, Timothy Krentz\IEEEauthorrefmark{1}, Gabor Karsai\IEEEauthorrefmark{1} and Srdjan Lukic\IEEEauthorrefmark{2}}
\IEEEauthorblockA{\IEEEauthorrefmark{1}Institute for Software-Integrated Systems, Vanderbilt University, Nashville, TN} 
\IEEEauthorblockA{\IEEEauthorrefmark{2}North Carolina State University, Raleigh, NC}}


\maketitle

\begin{abstract}
Executing distributed cyber-physical software processes on edge devices that maintains the resiliency of the overall system while adhering to resource constraints is quite a challenging trade-off to consider for developers. Current approaches do not solve this problem of deploying software components to devices in a way that satisfies different resilience requirements that can be encoded by developers at design time. This paper introduces a resilient deployment framework that can achieve that by accepting user-defined constraints to optimize redundancy or cost for a given application deployment. Experiments with a microgrid energy management application developed using a decentralized software platform show that the deployment configuration can play an important role in enhancing the resilience capabilities of distributed applications as well as reducing the resource demands on individual nodes even without modifying the control logic.
\end{abstract}


%
\IEEEpeerreviewmaketitle

\input{intro}
\input{relatedwork}
\input{appoverview}
\input{ftdesign}
\input{deploymentsolver}
\input{fttesting}
\input{conclusion}
\section*{Acknowledgment}
This work was funded in part by the Advanced Research Projects Agency-Energy (ARPA-E), U.S. Department of Energy, under Award Number DE-AR0000666. The views and opinions of the authors expressed herein do not necessarily state or reflect those of the US Government or any agency thereof.



\bibliographystyle{IEEEtran}
\bibliography{references,papers}



\end{document}

%% file: intro.tex
\section{Introduction}
\label{sec:intro}

The modern electric grid is a complex \textit{Cyber-Physical System (CPS)} \cite{Lee11introductionto} consisting of several software applications running simultaneously. They perform various control and regulatory functions associated with the system, for example, energy management \cite{boqtob2019microgrid}, protection \cite{li2020autonomous}, and automatic reconfiguration \cite{thakar2019system}. The various functions or roles required for these applications which are themselves distributed, can, in turn, be modularized into distributed computing processes or \texttt{actors}, like a relay actor for communicating with a relay between two sub-networks, a computation actor for calculating set points, etc. At run-time, the various actor processes are assigned to remote, embedded computing devices to execute their functions. This process is known as \textit{deployment}.


Existing deployment solutions do not allow for the consideration of the resilience requirements of such critical CPS and how they can be combined with the resource requirements of the software along with hardware limitations of the embedded nodes. This paper proposes a resilient deployment solver for calculating a valid deployment configuration that can be used to automate the deployment process such that incorporates resilience in terms of resilient operating constraints.

We implement our solution to the deployment and testing problems on a decentralized software platform for distributed software development called \textit{Resilient Information Architecture Platform for Smart Grids (RIAPS)} \cite{eisele2017riaps}, to leverage its inherent resilience features and ease of use. We test the deployment solution using a real-world application from the power systems domain and run it using using a virtual network emulator called \textit{Mininet} \cite{lantz2010network}.


Thus, the contributions of the paper are \textbf{(1)} A resilient deployment solver that generates a valid deployment configuration based on defined resilience constraints and \textbf{(2)} showcasing the effectiveness of the deployment in a practical energy management application. As a minor contribution, we also present the implementation of two fault tolerance patterns in the application using a RIAPS feature called \textit{groups}.
    
    
    


%% file: relatedwork.tex
\section{Related Work}
\label{sec:relatedwork}
 
Over the years, there has been substantial research on distributed systems deployment and synthesis that has, in turn, shaped our work. However, they are either restricted to only hardware reliability \cite{onishi2007solving}, \cite{glass2007reliability} , require fine-grained information \cite{gao2007component}, or require enforcing constraints in the design itself \cite{maticu2016automatic}.

 
 Cloud-fog computing architectures for Internet of things (IoT) applications also aim for the allocation of distributed processes or services to computing hardware while trying to optimize some criteria \cite{Smolka2022}, \cite{Roy2011}, \cite{Xia2018},\cite{Zhao2018}, \cite{Herrera2021}. However, no evaluation of resilience performance is performed after deployment. They mainly aim to reduce latency, response times, and manage node resources.   

Among other constraint-based works, Chariot \cite{Pradhan2018} employs a run-time self-reconfiguration algorithm for deploying IoT applications to edge devices, but it relies on commands from a continuously running manager component in all nodes and can not be used in design time. Zephyrus2 \cite{abraham2016zephyrus2} also introduces replication and separation constraints similar to ours and uses a specification language, but the platform is not designed to consider cyber-physical systems. The system architecture considered in \cite{8946711} assumes a fixed number of nodes, while \cite{manolios2010virtual} only considers resource allocation.

 

%% file: appoverview.tex
\section{Background}
\label{appoverview}

\begin{figure}[!ht]
\centering
\includegraphics[scale=0.3]{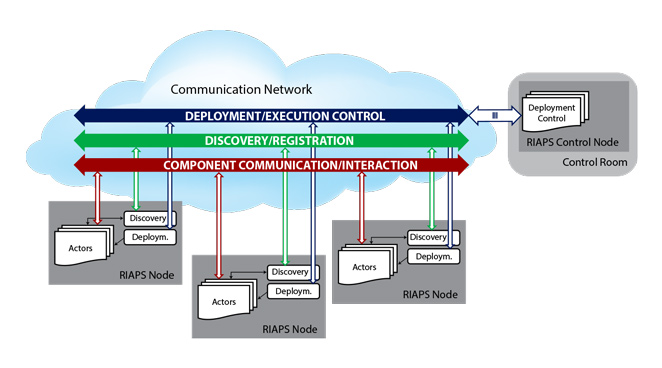}
\caption{RIAPS nodes, actors, components and  services}
\label{fig_riaps}
\end{figure}

\noindent \textbf{RIAPS Overview}: The \textit{Resilient Information Architecture Platform for Smart Grids (RIAPS)} \cite{eisele2017riaps} is an integrated and decentralized software framework that can be used for the development, deployment, and monitoring of distributed applications on remote hardware nodes, called RIAPS nodes. It supports a distributed computational model where a user can define \verb+actors+ containing a number of \verb+components+ that run different computations and can communicate with each other via message passing through interfaces called \verb+ports+. On top of it, it also provides a number of platform services for remote node control, coordination, fault monitoring, time synchronization and so on \cite{ghosh2019design}.  Figure \ref{fig_riaps} shows the RIAPS infrastructure. A RIAPS application is represented by an \texttt{app.riaps} file that describes the actor-component relationships along with the various message types and communication ports that each component uses. It also has an associated deployment file, \texttt{app.depl}, where the assignment of the actors to target computing nodes is specified.


\noindent \textbf{RIAPS Energy Management Application (REMApp)}: To show the effectiveness of our deployment solver and the resilience testing setup, we demonstrated prototyping a simple microgrid \cite{ton2012us} energy management application for a small office campus using RIAPS. It consisted of a building, an electric vehicle (EV) charging station, a battery energy storage system (BESS), and a centralized regulatory actor Aggregator, with functions as described in Table \ref{tab:remappactors}. A dispatch algorithm is designed to transfer available power to the loads and BESS while maintaining the battery limits.


The dispatch problem is formulated as an optimization problem and the constraints are described in Equation \ref{eq:optimize}. Here $x_{ij}$ denotes the power granted to load $i$ for time slot $j$, $K$ is the prediction horizon, $N$ number of electrical loads, ${P{grid}}_j$ denotes the total power available from the grid in time slot $j$ and ${P_{req}}_{ij}$ denotes the power required for unit $i$ at time slot $j$. $w_i$ is a weight assigned to each load which signifies its priority. Thus, the coordinator functions as a model predictive controller. Both the building and the charger actors consist of a predictive component, a manager component, and an interface device component. The predictive component incorporates a deep neural network based forecasting algorithm that is trained using prerecorded consumption data. We simulate building and charger loads using actual measurement data from a microgrid. We omit the implementation details of the REMApp application and the control algorithms, due to space constraints.
\begin{equation}
\label{eq:optimize}
\begin{aligned}
        \underset{x_{ij}}{min} \sum_{i=1}^{N}\sum_{j=1}^{K}{w_{i} \times 0.5 \times ({P_{req}}_{ij} - x_{ij}})^2 \\
    \sum_{i=1}^{N}{x_{ij}} \leq {P_{grid}}_j\  \forall j \\
    x_{ij} \geq 0 \\
    x_{ij} \leq {P_{req}}_{ij}
\end{aligned}
\end{equation}

\begin{figure}[!t]
\centering
\includegraphics[width=0.9\columnwidth]{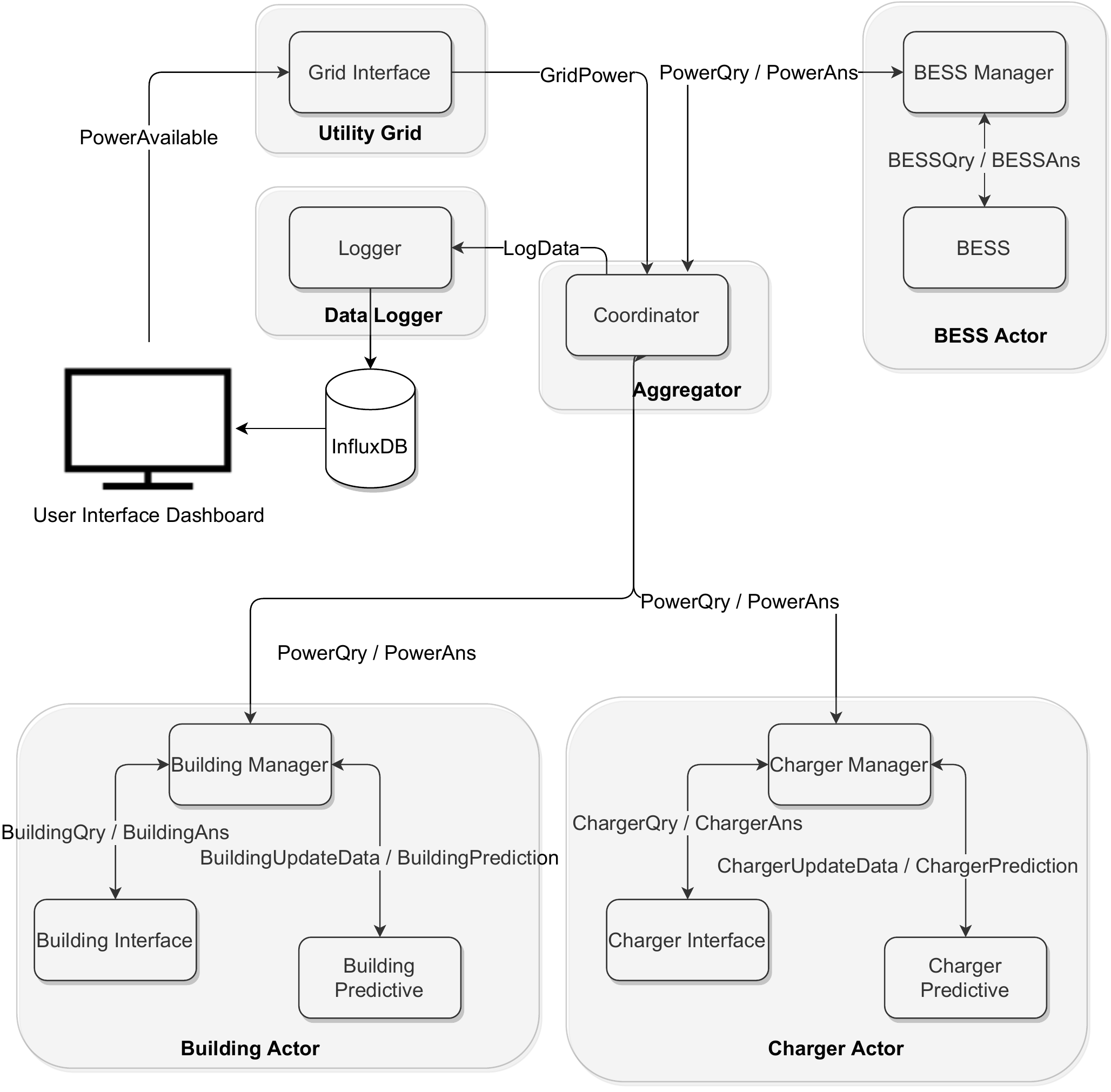}
\caption{RIAPS Energy Management Application. The translucent rectangles represent actors and the solid rectangles represent components. The arrows indicate messages exchanged. This is the logical architecture prior to including any redundancy for fault tolerance.}
\label{fig_remapp}
\end{figure}

Apart from these actors, we also implement a Data Logger actor with a Logger component to log measurement and event-related data and also run a remote dashboard for visualizing the results. The application model is shown in Figure \ref{fig_remapp}.

\input{tables/remappactors}

%% file: tables/remappactors.tex
\begin{table}[!t]
\centering
\begin{tabular}{|p{0.2\columnwidth}|p{0.6\columnwidth}|}
\hline
\textbf{Actors}                & \multicolumn{1}{p{0.7\columnwidth}|}{\textbf{Functions}}                                                             \\ \hline
\multirow{3}{*}{\begin{tabular}[c]{@{}l@{}}Building \\ and Charger \\ actors\end{tabular}} & \multicolumn{1}{p{0.7\columnwidth}|}{\textbf{1.}~Get power consumption readings from sensors}                                   \\
                                      & \textbf{2.}~Estimate power demands for a future time horizon (e.g. a day or 12 hours)                           \\
                                      & \textbf{3.}~Send power request to the coordinator, receive setpoint commands from the coordinator               \\ \hline
Grid actor &
  \multicolumn{1}{p{0.7\columnwidth}|}{\textbf{1.}~In grid-connected mode, send information to the coordinator about the power available from the main grid for the time horizon} \\ \hline
\multirow{2}{*}{BESS actor} &
  \multicolumn{1}{p{0.7\columnwidth}|}{\textbf{1.}~Act as an energy buffer when the microgrid operates in islanded mode or if the power demand can not be met by the grid}  \\ \hline
\multirow{4}{*}{Aggregator}          & \multicolumn{1}{p{0.7\columnwidth}|}                  {\textbf{1.}~Produce an optimal allocation strategy for each future time step based on the power available} \\
 &
  \textbf{2.}~Based on the power available and BESS state of charge, send discharging/ charging commands to it. \\
                                      & \textbf{3.}~Provide allocated power to the loads                                                                   \\ \hline
\end{tabular}
\medskip
\caption{Actors and their functions in REMApp}
\label{tab:remappactors}
\end{table}

%% file: ftdesign.tex
\label{sec:ftdesign}


\noindent \textbf{RIAPS Groups}:  In RIAPS, A \verb+group+ defines a virtual connection bus between components on the actual physical network. It implements a pub-sub pattern among members providing features like intra- group communication, dynamic joining and leaving, leader election (based on Raft \cite{ongaro2014search}), and member voting on a value or an action. While the primary application of groups is to create a hierarchical architecture within a system for multilevel coordination and control, we demonstrate here that some of those features can be leveraged to implement robust fault tolerance design patterns.

Considering the REMApp application, its principal performance and safety specifications can be described as:

\noindent \textbf{1.} The system should not dispatch more power than is available at the current time step.

\noindent \textbf{2.} The power allocated to the loads should be close to the predicted demand, provided the first condition is not violated.

For the final application design, we consider two kinds of threats for the application: \textbf{(1)} \textit{Node drop out} caused by either a communication fault or a hardware crash and \textbf{(2)} \textit{False Data Injection security Attack (FDIA)} occurring at any of the sensors in the building or the EV charger leading to erroneous power allocation.

In order to make the system resilient to these two types of threats, we instantiate redundant copies of actors and introduce two types of group behavior protocols for the components within those actors, as described next.

\noindent \textbf{Redundancy groups}: Provides an automated failover mechanism.  The behavior is the following: Elect a leader, detect if leader is out, and then replace with a new leader.

\noindent \textbf{Consensus groups}: It is a distributed agreement pattern on top of leader election. Each member initiates a vote on a value; if leader's vote is successful, then it uses its own value. Otherwise, it selects a group member with a verified vote to send their value.

For the REMApp application, we apply the redundancy groups pattern to the coordinator component and the consensus groups pattern to the building, EV charger and BESS manager components.

%% file: deploymentsolver.tex
\section{Resilient Deployment Solver}
\label{sec:deploymentsolver}


The deployment solver was implemented in Python and consists of four parts: a deployment specification language \textit{dspec} using which users can define fault tolerance as well as resource specifications, parsed using a model parser; a constraint solver implemented using Z3 \cite{Moura2008} that converts them into constraints and computes a valid deployment matrix satisfying them; and finally a deployment generator that converts the solution into a RIAPS deployment file. It takes as input a RIAPS model file \texttt{app.riaps}, a deployment specification file \texttt{app.dspec}, a hardware configuration file \texttt{hardware-spec.conf} for resource limits and produces as output a RIAPS deployment file \texttt{app.depl}.

\subsection{Constraint formulation}
\label{sub:constraints}

Before formalizing the constraints used in the solver, some basic definitions are required.

\noindent \textbf{Actor}: An actor is a tuple consisting of an identifier $ID$, redundancy $D$, host dependency $H$ and a set of resource constraints $R$. $A_j := \langle ID, D, H, R \rangle$.

\noindent \textbf{Node}: A node is a tuple consisting of an identifier $ID$ and a set of resource constraints $R$. $Node_i := \langle ID, R \rangle$

\noindent \textbf{Redundancy}: It is an integer that implies the number of copies of an actor that needs to be deployed. $D(A_j) \in \mathbb{N}$

\noindent \textbf{Host dependency}: The host dependency for an actor is a mapping between an actor and a node that the actor must be deployed to. $H(A_j) = Node_i.ID$. This captures the scenario of actors tied to a particular host, say, for example edge located devices that must be accessed from that particular location.

\noindent \textbf{Colocation condition}: It is a set of actors that must be deployed to the same node. $C = \{ A_j | j \in \mathbb{N} \}$.

\noindent \textbf{Separation condition}: It is the opposite of colocation. It is a set of actors that can not be placed on the same node. $S = \{ A_j | j \in \mathbb{N} \wedge (A_j, A_k \in S \implies A_j, A_k \notin C.\ \forall i \neq j ) \}$

\noindent \textbf{Resource limits}: They are a set $R()$ specifying the worst case resource usage metrics defined for the actors that must be limited by the hardware resource specifications in case of nodes, including, CPU utilization ($cpu$, $cpu_{max}$ for worst-case) per core up to total $cores$, expressed as a percentage over $interval$, memory ($mem$) expressed in MB, and disk space ($spc$) expressed in MB. For network bandwidth two quantities are used, an average rate, ($rate$) and a maximum ceiling ($ceil$), expressed in kilobits per second (kbps).

The various resource limits for the hardware are specified using a configuration file with the name of the hardware device acting as the key. An example is shown in Figure \ref{fig:hwspec}. The network specifications for the node are taken from the RIAPS configuration file included with RIAPS installation, which contains values specified for the network interface card rate (\textit{NIC\_RATE}) and ceiling (\textit{NIC\_CEIL}). The worst case resource usage for the various actors are specified in the RIAPS model under the individual actor definitions. An example is shown in figure \ref{fig:actorspec}. 

\begin{figure}
    \centering
\begin{lstlisting}[language=riaps]
# Hardware Specifications
#Beaglebone Black
[bbb]
# no. of CPU cores
cores = 1
max_cpu = 0.7 # percentage
# Internal memory (MB)
mem = 512
max_mem = 0.7
# disk space (MB)
spc = 4096
max_spc = 0.7
\end{lstlisting}
\caption{Hardware specification file. The name of the hardware device is used as the key. Thus, the same file can be used to deploy to a network of heterogeneous devices.}
\label{fig:hwspec}
\end{figure}

\begin{figure}
    \centering
\begin{lstlisting}[language=riaps]
actor Aggregator(configfile, id, grptype) {
	   uses {
			cpu max 35 % over 1;		// cpu limit over 1 second
			mem 200 mb;					// Mem limit
			space 2048 mb;				// File space limit
			net rate 40 kbps ceil 60 kbps; // Net limits
		}
   ...
 }
\end{lstlisting}
\caption{The resource limits placed on the actor can be defined in the RIAPS model under the \texttt{uses} block.}
\label{fig:actorspec}
\end{figure}

The final deployment solution space is encoded as a 2D deployment matrix where each element is a binary variable, $X$ defined as equation \ref{eq:1}.

\begin{equation}
    \label{eq:1}
    X_{i,j} = \begin{cases}
    1, & \text{if $A_j$ assigned to $Node_i$}. \\
    0, & \text{otherwise}.
    \end{cases}
\end{equation}

Based on these definitions the different constraints used by the solver can be formulated as:

\noindent \textbf{Redundancy}
\begin{equation}
    \forall j ( \sum_i X_{i,j} = D(A_j))
\end{equation}
\noindent \textbf{Host-actor dependency}
\begin{equation}
    \forall j (Node_i \in H(A_j)) \implies (\forall p \neq i,\ X_{p,j} = 0)
\end{equation}
\noindent \textbf{Colocation constraints}
\begin{equation}
    \forall A_m, A_n \in C, (X_{i,m} = 1) \implies (X_{i,n} = 1)
\end{equation}

\noindent \textbf{Separation constraints}
\begin{equation}
    \forall A_m, A_n \in S, (X_{i,m} = 1) \implies (X_{i,n} = 0)
\end{equation}

\noindent \textbf{CPU}
\begin{equation}
\begin{split}
    \forall i \sum_j (X_{i,j} \times cpu(A_j))/100*interval \\ < cpu_{max}(Node_i)/100*cores*interval
    \end{split}
\end{equation}

\noindent \textbf{Memory}
\begin{equation}
    \forall i \sum_j (X_{i,j} \times mem(A_j)) < mem(Node_i)
\end{equation}

\noindent \textbf{Disk space}
\begin{equation}
    \forall i \sum_j (X_{i,j} \times spc(A_j)) < spc(Node_i)
\end{equation}

\noindent \textbf{Network bandwidth}
\begin{equation}
\begin{split}
 \forall i \sum_j (X_{i,j} \times rate(A_j))+Max(ceil_{j}-rate_{j}) \\ < 0.95*NIC\_RATE(Node_i)   
\end{split}
\end{equation}


We applied the solver to the REMApp model to validate its correctness, as well as evaluate its performance in terms of scalability and timing. 
\begin{figure}
    \centering
\begin{lstlisting}[language=riaps]
app REMApp {
	Aggregator copies 2;
	colocate (UtilityGrid,DataLogger);
	separate (BESSActor, BuildingActor, ChargerActor);
	deploy (UtilityGrid) on (h1);
	// use limits for bbb on all;
	}
\end{lstlisting}
\caption{REMApp.dspec file. The last line (commented here) is used for resource constraints.}
\label{fig:dspec}
\end{figure}

\noindent \textbf{Basic Solver}: We use the specifications in Figure \ref{fig:dspec} to run the solver for 12 nodes. The solution produced by the solver deploys $1$ copy of the \textit{DataLogger} and \textit{UtilityGrid} actors on host \textit{h1}, $1$ copy of the \textit{Aggregator} actor each on hosts \textit{h5} and \textit{h8}, $1$ copy of \textit{BESSActor} on host \textit{h5}, $1$ copy of \textit{ChargerActor} on host \textit{h7} and $1$ copy of \textit{BuildingActor} on host \textit{h10}. Thus, it utilizes only 5 out of the 12 nodes.


\noindent \textbf{Maximum Redundancy}: By tweaking the solver configuration, it is possible to search for a solution that maximizes the copies of actors given a number of nodes. In this case, there were $12$ copies of \textit{Aggregator}, $1$ per node; $3$ copies of the \textit{BESS} actor on \textit{h7, h8, h9}; $3$ copies of \textit{BuildingActor} on \textit{h1, h5, h11}; and $6$ copies of \textit{ChargerActor} on \textit{h2, h3, h4, h6, h10, h12}. Since the \textit{DataLogger} and \textit{UtilityGrid} actors are co-located and tied to one host, there was only $1$ copy of them on \textit{h1}.

\noindent \textbf{Minimum Cost}: The solver can also be configured to solve for a deployment configuration that uses the minimum number of nodes for the given constraints. 

This setting was used to generate the deployment for our test run by modifying the specifications of figure \ref{fig:dspec} in the following way: 

\begin{itemize}
    \item Include 3 copies of the building, charger and BESS actors to tolerate a single actor failure due to FDIA attack as per the $2n+1$ redundancy rule \cite{castro1999practical}. 
    \item Separate the charger and the aggregator actor. This was done purely for experimental purposes to demonstrate the effects of the faults on the aggregator and the charger independently.
\end{itemize}
The solution indicates a minimum of $9$ nodes to deploy the application as shown in table \ref{tab:9node}. 

\input{tables/9node}



The solver was tested on an Intel desktop PC with 8 GB RAM, on a number of problems and the computation time stayed below 5 seconds for up to 100 nodes and below 9 seconds for up to 1600 assertions added to the solver. This is comparable to values reported in \cite{Pradhan2018}, which reports an average of $1.74$ seconds and a maximum of $48$ seconds.

%% file: tables/9node.tex
\begin{table}[!t]
\centering
\begin{tabular}{p{0.02\columnwidth}|p{0.1\columnwidth}|p{0.1\columnwidth}|p{0.1\columnwidth}|p{0.1\columnwidth}|p{0.1\columnwidth}|p{0.1\columnwidth}|}
\cline{2-7}
                                    & Aggre- gator & BESS Actor & Building Actor & Charger Actor & Data Logger & Utility Grid \\ \hline
\multicolumn{1}{|l|}{h1} & 0          & 0         & 1             & 0            & 1          & 1           \\ \hline
\multicolumn{1}{|l|}{h2} & 1          & 0         & 1             & 0            & 0          & 0           \\ \hline
\multicolumn{1}{|l|}{h3} & 0          & 1         & 0             & 0            & 0          & 0           \\ \hline
\multicolumn{1}{|l|}{h4} & 0          & 1         & 0             & 0            & 0          & 0           \\ \hline
\multicolumn{1}{|l|}{h5} & 0          & 0         & 0             & 1            & 0          & 0           \\ \hline
\multicolumn{1}{|l|}{h6} & 0          & 1         & 0             & 0            & 0          & 0           \\ \hline
\multicolumn{1}{|l|}{h7} & 0          & 0         & 0             & 1            & 0          & 0           \\ \hline
\multicolumn{1}{|l|}{h8} & 1          & 0         & 1             & 0            & 0          & 0           \\ \hline
\multicolumn{1}{|l|}{h9} & 0          & 0         & 0             & 1            & 0          & 0           \\ \hline
\end{tabular}
\medskip
\caption{Solver solution for the fault tolerance testing configuration}
\label{tab:9node}
\end{table}

%% file: fttesting.tex
\section{Fault Tolerance Testing}
\label{sec:fttesting}


We tested the application using a virtual network emulator Mininet \cite{lantz2010network}. Mininet allows for rapid prototyping of large networks on a single computer. It also provides a command-line interface (CLI) that can be used to control link parameters, bring links up or down, and send commands to hosts.

\subsection{Experimental Results}

\begin{figure}[!t]
    \centering
    \includegraphics[scale=0.25]{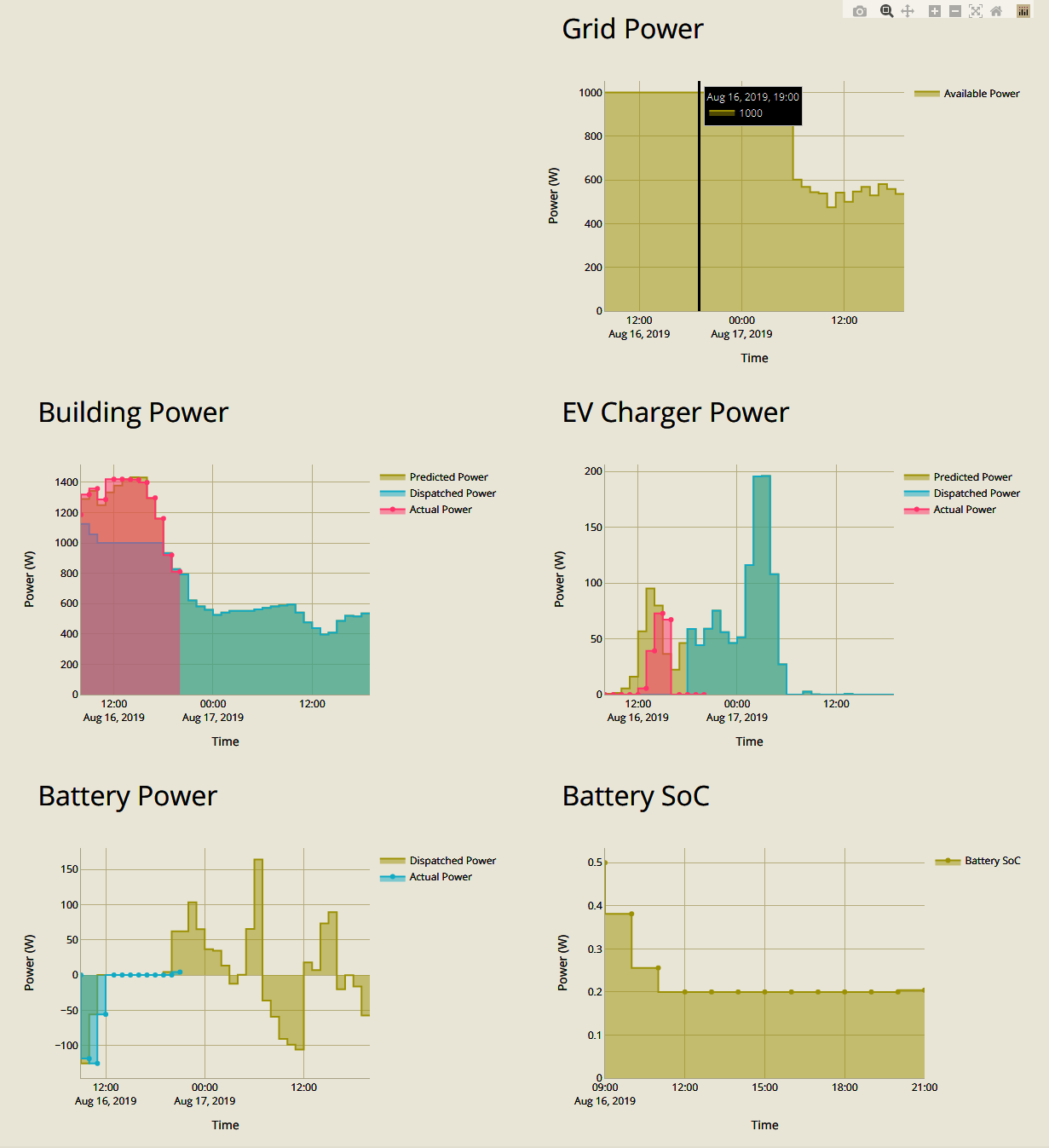}
    \caption{Snapshot of run 3 showing the energy profiles of the different loads and the power from the grid. For the loads, The light green curve represents the predicted power, the light blue curve represents the power dispatched by the coordinator. The green line is the measurment data.}
    \label{fig:remappgui}
\end{figure}

\label{sec:experiment}
A 60 minute resilience test was designed and tested with the deployment generated in Table \ref{tab:9node} for the constraints described in Section \ref{sec:deploymentsolver}.

\noindent \textbf{1.} At $t_0$, start test.

\noindent \textbf{2.}  At $t_0 + 20$, bring down host h2 (Aggregator).

\noindent \textbf{3.}  At $t_0 + 40$, start the FDIA attack on host h7 (EV Charger).

\noindent \textbf{4.}  At $t_0 + 55$ stop the attack.

\noindent \textbf{5.}  At $t_0 + 60$, end test.



The experiments were repeated three times. For two of the three runs, the first fault targeted a leader node (leader election is dynamic). This triggered a new leader election, which took about $1$s on average. The average time interval between the start of a round of voting and the obtaining of its result was $9$ ms before and $19$ ms after \textit{FDIA}. Both of these times are negligible when compared to the controller time-step of $15$ minutes real time that was used in the application.

Figure \ref{fig:remappgui} shows how the system performed during run 3 of the experiment. The black vertical line indicates the end time of the simulation. The right of that line indicates the future time horizon. Validating the plots in light of the performance criteria discussed in section \ref{sec:ftdesign}, it can be observed that \textit{Condiiton 1} was satisfied, indicated in the pre-fault period (left of Figure \ref{fig:remappgui}) when the power dispatched to both loads was less than the predicted or requested power. This was because the available power was not adequate, even after the battery was discharged, as seen in the state-of-charge (SoC) plot. This condition was also satisfied during the post-fault period where the power dispatched for the building was capped to 1000 KW, while the chargers encountered almost total load curtailment. This was due to the fact the the battery SoC was down to its minimum specification of $0.2$ and hence unable to discharge. \textit{Condition 2} was also satisfied as observed in the plots that the allocated power closely followed the predicted power and only deviated in case the available power was insufficient, showing that the application functionality was resilient to both threats.

\subsection{Deployment with Network Bandwidth Constraints}

This section demonstrates how the solver evaluated the deployment configuration taking into account the resource limiting constraints for the hardware devices and resource requirements for the distributed application actors. Network bandwidth constraints were used to conduct the experiments since the data can be easily captured and isolated for each mininet host. 


\input{tables/res12node}

\begin{figure}[!h]
    \centering
    \begin{subfigure}[h]{\columnwidth}
    \centering
    \includegraphics[width=\columnwidth, height=4cm]{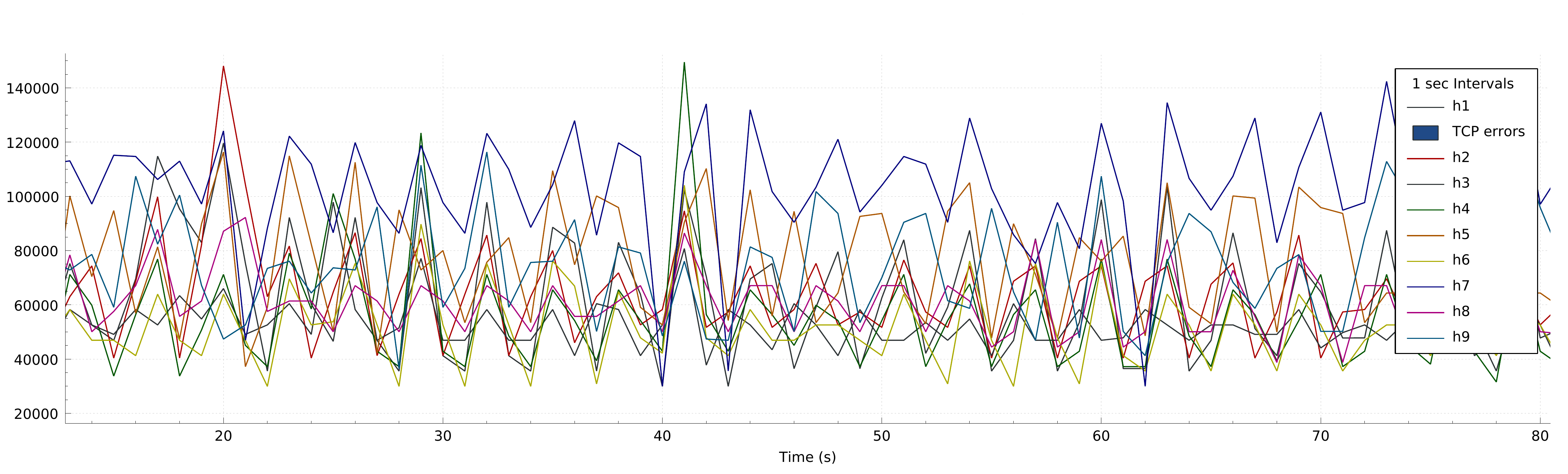}
    \caption{Network traffic for deployment without using resource constraints}
    \label{fig:9nodenores}
    \end{subfigure}
    \hfill
    \begin{subfigure}[h]{\columnwidth}
    \centering
    \includegraphics[width=\columnwidth, height=4cm]{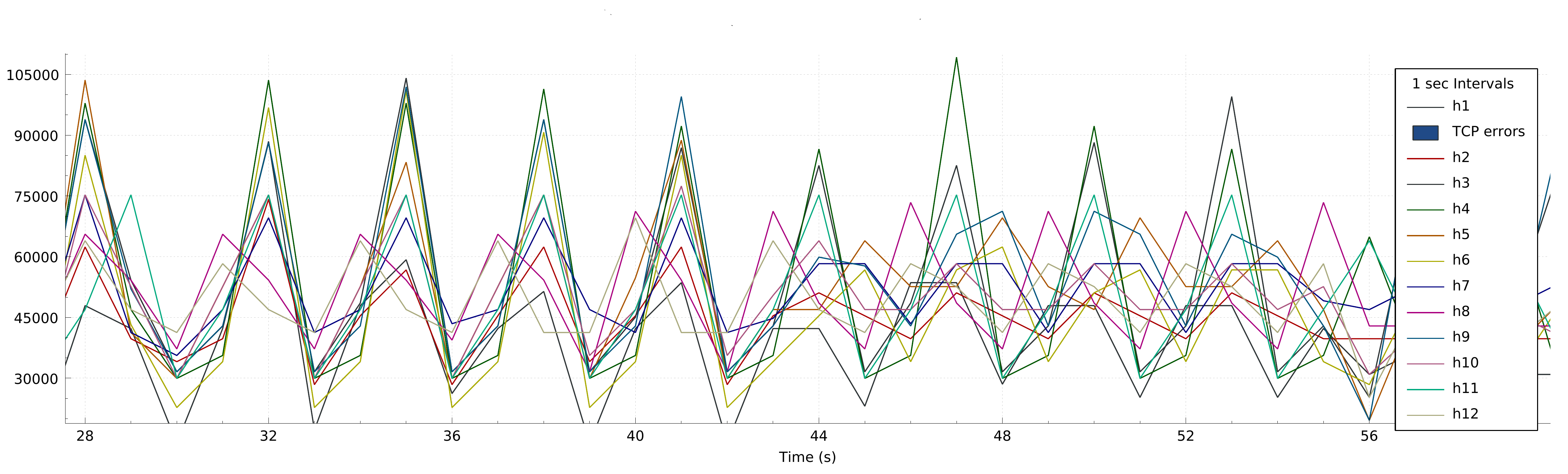}
    \caption{Network traffic for deployment after using resource constraints}
    \label{fig:12noderes}
    \end{subfigure}
    \caption{Network traffic generated by the hosts while running REMApp. The vertical axis represents data in bits.}
    \label{fig:nwexp}
\end{figure}

The network interface specifications chosen to run the experiment were $NIC\_RATE = 118 kbps$ and $NIC\_CEIL = 131 kbps$ (Sec \ref{sec:deploymentsolver}: Resource limits). The network limits the REMApp actors application were specified as $rate=40 kbps$ and $ceil = 60 kbps$ for each actor. These values were chosen in this experiment mainly for demonstration purposes. In a production deployment, typically, each actor will have different values which need to be calculated by prior bench marking or building faithful models and simulation. The solver was run using the same \textit{dspec} file described in figure \ref{fig:dspec} but with the last line uncommented, allowing resource limits. The solution obtained is shown in Table \ref{tab:res12node}, which uses 3 additional nodes compared to Table \ref{tab:9node} without using resource limits.


Figure \ref{fig:nwexp} shows the \texttt{tcp} network traffic for each host (node) while running the application captured over $1$ second intervals. Figure \ref{fig:9nodenores} corresponds to the configuration of Table \ref{tab:9node}, while figure \ref{fig:12noderes} corresponds to Table \ref{tab:res12node}). The \texttt{tcp} packets include the application messages sent via RIAPS along with some overhead. From the graphs, it can be seen that the average bandwidth consumed by the hosts is $84k$ bits per second in Figure \ref{fig:9nodenores} and $71k$ bits per second in Figure \ref{fig:12noderes}, indicating a reduction of $15 \%$. Looking at the maximum bandwidth consumed, it can be seen that it remained almost entirely within $100k$ bits for all hosts, when solved with network limits, with a few overshoots. However, they were within the $NIC\_RATE$ of $118 kbps$ set for the calculations. On the other hand, the bandwidth consumed by the host \textit{h1} regularly exceeded $120 kbps$ when deployed without using limits.

%% file: tables/res12node.tex
\begin{table}[!t]
\centering
\begin{tabular}{p{0.01\columnwidth}|p{0.1\columnwidth}|p{0.1\columnwidth}|p{0.1\columnwidth}|p{0.1\columnwidth}|p{0.1\columnwidth}|p{0.1\columnwidth}|}
\cline{2-7}
                                     & Aggre- gator & BESS Actor & Building Actor & Charger Actor & Data Logger & Utility Grid \\ \hline
\multicolumn{1}{|l|}{h1}  & 0          & 0         & 0             & 0            & 1          & 1           \\ \hline
\multicolumn{1}{|l|}{h2}  & 1          & 0         & 0             & 0            & 0          & 0           \\ \hline
\multicolumn{1}{|l|}{h3}  & 0          & 0         & 0             & 1            & 0          & 0           \\ \hline
\multicolumn{1}{|l|}{h4}  & 0          & 0         & 1             & 0            & 0          & 0           \\ \hline
\multicolumn{1}{|l|}{h5}  & 0          & 0        & 0            & 1            & 0          & 0           \\ \hline
\multicolumn{1}{|l|}{h6}  & 1          & 0         & 0             & 0            & 0          & 0           \\ \hline
\multicolumn{1}{|l|}{h7}  & 0          & 1         & 0             & 1            & 0          & 0           \\ \hline
\multicolumn{1}{|l|}{h8}  & 0          & 1         & 0             & 0            & 0          & 0           \\ \hline
\multicolumn{1}{|l|}{h9}  & 0          & 0         & 1             & 0            & 0          & 0           \\ \hline
\multicolumn{1}{|l|}{h10} & 0          & 0         & 0             & 1            & 0          & 0           \\ \hline
\multicolumn{1}{|l|}{h11} & 0          & 1         & 0             & 0            & 0          & 0           \\ \hline
\multicolumn{1}{|l|}{h12} & 0          & 0         & 1             & 0            & 0          & 0           \\ \hline
\end{tabular}
\medskip
\caption{Solver solution for 12 nodes with network limits}
\label{tab:res12node}
\end{table}

%% file: conclusion.tex
\section{Conclusion and Future Work}
\label{sec:conclusion}

This paper introduced a design time framework for prototyping and deploying resilient distributed applications for cyber-physical systems such as the Smart Grid. The resilient deployment solver maps distributed application entities or actors to remote nodes while adhering to both hardware and resilience specifications. The deployment was tested using a real world application of a microgrid energy management system and successfully demonstrated how the deployment decision can affect the behavior of a distributed application under faults and can also help maintain the resource limitations imposed by the hosting platform.

Future work would include incorporating latency and timing constraints like \cite{8946711}, and implementing more fault tolerance patterns.